\documentclass[aps,prl,preprint,superscriptaddress]{revtex4-2}

\usepackage{graphicx}
\usepackage{color}

\begin{document}



\title{Pressure induced quenching of planar rattling in Cu$_{10}$Zn$_{2}$Sb$_{4}$S$_{13}$ studied by specific-heat and x-ray diffraction measurements}



\author{Kazunori Umeo}
\email[e-mail address: ]{kumeo@hiroshima-u.ac.jp}
\affiliation{Department of Low Temperature Experiment, Integrated Experimental Support / Research Division, N-BARD, Hiroshima University, Higashi-Hiroshima, 739-8526, Japan}
\author{Koichiro Suekuni}
\affiliation{Department of Applied Science for Electronics and Materials, Interdisciplinary Graduate School of Engineering Sciences, Kyushu University, Fukuoka 816-8580, Japan}
\author{Toshiro Takabatake}
\affiliation{Graduate School of Advanced Science and Engineering, Hiroshima University, Higashi-Hiroshima 739-0826, Japan}
\author{Eiji Nishibori}
\affiliation{Department of Physics, Faculty of Pure and Applied Sciences, and Tsukuba Research Center for Energy Materials Science (TREMS), University of Tsukuba, 1-1-1 Tennodai Tsukuba, Ibaraki, 305-8571, Japan}


\date{\today}

\begin{abstract}

We have studied the pressure effect on the rattling of tetrahedrite 
Cu$_{10}$Zn$_{2}$Sb$_{4}$S$_{13\,}$(CZSS) and type-I clathrate 
Ba$_{8}$Ga$_{16}$Sn$_{30\,}$(BGS) by specific heat and 
x-ray diffraction measurements. By applying pressure $P$, the rattling energy for CZSS 
initially decreases and steeply increases for $P \, \textgreater \, 1$ GPa. By 
contrast, the energy for BGS increases monotonically with $P$ up to 6.5 
GPa. An analysis of the pressure dependent specific heat and x-ray 
diffraction indicates that the out-of-plane rattling of the Cu atoms in the 
S$_{3}$ triangle of CZSS originates from the chemical pressure, unlike the 
rattling of the Ba ions among off-center sites in an oversized cage of BGS. 
The rattling in CZSS ceases upon further increasing $P$ above 2 GPa, suggesting 
that Cu atoms escape away from the S$_{3}$ triangle plane.

\end{abstract}


\maketitle


\newpage

Localized vibrational states of atoms in solids are attracting much 
attention in the fields of solid state physics and materials science \cite{VK01,TT02}. 
Such states are realized in compounds with a three-dimensional network of 
polyhedral cages including one guest atom. When the cage is oversized 
compared to the guest atom, so called ``rattling modes'' can be realized \cite{GA03}. 
This concept was introduced by Slack to describe the situation where the guest 
atom moves among two or more semistable positions and no long-range 
correlation exists between their positions or orientations \cite{GA03}. The rattling 
modes give rise to unusual physical properties: a strong suppression of the 
lattice thermal conductivity $\kappa_{\rm L}$ in some type-I clathrates 
$X_{8}$Ga$_{16}$Ge$_{30}$ ($X =$ Eu, Sr) and Ba$_{8}$Ga$_{16}$Sn$_{30}$ 
(BGS) \cite{BC04,MA05}, and superconductivity in $\beta $-pyrochlore oxides $A$Os$_{2}$O$_{6\, }$($A =$ K, Rb, Cs) \cite{YN07} and VAl$_{10.1}$ \cite{DJ08}. Because low $\kappa_{\rm L}$ is one of 
the requisites for thermoelectric materials, a number of experimental and 
theoretical works have been conducted to clarify how the rattling scatters 
acoustic phonons \cite{TT02,TT09}. 

Rattling vibrations in caged compounds depend on the difference between the 
cage size and the guest size  \cite{TT02}. When the cage size is reduced by applying 
pressure or by substituting the cage atoms with smaller ones, it is expected 
that the rattling energy is increased and the anharmonicity in the 
vibrational mode is weakened. Among clathrates, one of the well-studied 
examples is BGS, where the Ba guest atom exhibits so-called off-center 
rattling in a tetrakaidekahedral cage consisting of Ga and Sn atoms \cite{MA05,KS10}. 
When the cage of BGS is pressurized up to 6 GPa at room temperature, the 
guest atom moves from an off-center to on-center position and the energy of the 
rattling vibration increases, as found by high-pressure Raman scattering 
experiments \cite{TS11}. When the cage size of Sr$_{8}$Ga$_{16}$Ge$_{30}$ is 
decreased by the substitution of Si for Ge, the rattling energy is also 
enhanced, as found by specific-heat measurements \cite{KS12}. 

Recently, rattling phenomena have been found even in non-caged compounds in 
which a rattling atom has a planar coordination. For example, large 
amplitude out-ofplane motion, which is called ``planar rattling'' \cite{CH13}, 
occurs for Bi atoms in LaOBiS$_{2x}$Se$_{x\, }$\cite{AN14,YM15} and Cu atoms in 
tetrahedrite Cu$_{12-x}Tr_{x}$Sb$_{4}$S$_{13}$ ($Tr=$ Mn, Fe, Co, Ni, Cu, 
and Zn) \cite{XL16,KS17}. This new type of rattling has extended the classes of 
materials in search of highperformance thermoelectric materials with small 
$\kappa_{\rm L}$ \cite{AN14,XL16,KS17,KS18,YB19}.

The tetrahedrite crystallizes in a cubic structure with space group $I\bar{4}3m$ as 
shown in the inset of Fig. 1 \cite{AP20}. Note that the phase transition at 85 K 
accompanied by a structural transformation in pure tetrahedrite 
Cu$_{12}$Sb$_{4}$S$_{13}$ \cite{KS18,SK21,HI22,AF23,DI24,NG25,VR26,SO27} is avoided by partial substitution of 
$Tr$ for Cu1 occupying a tetrahedron composed of four S1 atoms \cite{KS18,AF23}. The 
Cu2 atom is located in a triangle composed of two S1 atoms and one S2 atom, 
and has a large atomic displacement parameter (ADP) out of the S$_{3}$ 
triangle and toward the Sb atoms for the cubic structure \cite{KS17}. 
The low-energy quasi-localized vibrational mode of Cu2 has been revealed by neutron scattering measurements \cite{KS18,AF23,KS28}. A systematic study of the crystal structure and phonon structure for Cu$_{10}$Zn$_{2}$Sb$_{4}$S$_{13}$ (CZSS), 
Cu$_{12}$Sb$_{4}$S$_{13}$ (CSS), Cu$_{10}$Zn$_{2}$As$_{4}$S$_{13}$, and 
Cu$_{12}$As$_{4}$S$_{13}$ \cite{KS28} demonstrated that the out-of-plane ADP for the Cu2 atom increases and simultaneously the characteristic energy of the out-of-plane vibration is reduced with decreasing the area of the S3 triangle. 
These results support the model that the rattling 
stems from the chemical pressure inherent in the triangle to squeeze the Cu2 
atom out of the plane. On the other hand, based on the first-principles 
calculation, Lai \textit{et al}. \cite{WL29} have claimed that the rattling of the Cu2 atom in CSS is 
attributed to the creation of a covalent-type bonding between Cu and Sb 
through lone pairs of the trivalent Sb ion. Therefore, the driving mechanism 
of the rattling of the Cu2 atom in tetrahedrite has thus far remained an open question. If we measure the pressure effect on the rattling energy while 
keeping the chemical species (i.e. Sb or As) around the Cu2 atom intact, the 
dominant factor for the occurrence of rattling could be extracted. More 
specifically, it could be clarified how the rattling mode depends on the 
area of the S$_{3}$ triangle and the Cu2-Sb distance.

The rattling mode in caged compounds manifests itself in a broad peak of the 
specific heat $C$ divided by $T^{3}$, whose temperature is approximately 1/5 of 
the characteristic energy of rattling \cite{LG30}. For both tetrahedrite 
CZSS and type-I clathrate BGS, a peak in $C$/$T^{3}$ appears at 4 K, indicating 
the characteristic energy of the rattling mode to be 20 K \cite{MA05,KS10,KS31}. 

In this Rapid Communication, we have compared the pressure dependences of the rattling modes for 
CZSS and BGS by specific heat measurements to shed light on the mechanism 
of the rattling in CZSS. We have found that application of pressures $P$ up to 0.7 GPa 
reduces the rattling energy of CZSS but increases monotonically that of BGS. 
This opposite changes strongly indicate that the origin of anharmonicity in 
the rattling is different between CZSS with the planar coordination and BGS 
with the oversized cage. Upon further increasing $P$ above 2.4 GPa, the 
rattling contribution to $C$/$T^{3}$ for CZSS is fully suppressed, while the 
vibrational density of states in BGS at $P = 6$ GPa remains more than 50{\%} 
of that at $P = 0$. The cessation of rattling for $P \, \textgreater \, 2.4$ GPa causes a 
profound change in the vibration of the Cu atom from anharmonic rattling to a rather 
harmonic vibration in the local minimum of the potential out of the S$_{3}$ 
triangle. The splitting of the Cu site is corroborated with synchrotron 
x-day diffraction measurements under pressure. 

Samples of polycrystalline Cu$_{10}$Zn$_{2}$Sb$_{4}$S$_{13}$ (CZSS) and 
single-crystalline Ba$_{8}$Ga$_{16}$Sn$_{30}$ (BGS) were synthesized in the 
manners as described elsewhere \cite{MA05,KS31}. The measurement of the specific heat 
$C$ was performed by the ac method in pressure and temperature ranges of $P \, \le \, 
 6.5$ GPa and $0.5 \, \textless \, T \, \textless \, 10$ K, respectively. Thereby, a 
Bridgman anvil cell was installed in a $^{3}$He cryostat \cite{KU32}. The sample 
was wrapped in indium foil, which played the role of a pressure transmitting 
medium. The wrapped sample was packed in the cell by a gasket made of 
Cu--Be. The weights of the indium foil and gasket were 8.05 and 9.11 mg for 
CZSS (2.60 mg), and 9.52 and 9.09 mg for BGS (4.57 mg), respectively. Two 
chip resistors were used as the thermometer and the heater which were 
mounted on the outer flange of the gasket. Because the thermometer is free 
from pressure, it was not necessary to be calibrated under different 
pressures. The pressure was estimated by the pressure dependence of the 
superconducting transition temperature of the In foil. The details of the 
experimental set-up were described in Ref. [31]. The pressure dependence of the
crystal structure of CZSS was examined at room temperature by synchrotron 
powder x-ray diffraction at the SPring-8, BL02B1 beamline. A pressure of up 
to 1.9 GPa was applied using a diamond anvil cell. By analyzing the 
diffraction patterns, we determined the atomic position and interatomic 
distances. 

Figure 1 shows the temperature dependence of $C$/$T^{3}$ for CZSS under various 
$P$ up to 3.1 GPa. The data at $P = 0$ with a broad maximum at 4 K agree with 
the previous data (open circles) measured by a Quantum Design physical property measurement system (PPMS) \cite{KS31}. 
With increasing pressure up to 0.7 GPa, the value of $C$/$T^{3}$ at around 2 K 
increases as a result of the shift of the broad maximum of $C$/$T^{3}$ to low 
temperatures. Above 1 GPa, the $C$/$T^{3}$ values at $T \, \textgreater \, 2$ K decrease 
gradually, and for $P \ge  2.4$ GPa, reach the value of the Debye specific 
heat with a Debye temperature of 242 K as shown by the dashed line \cite{KS31}. 
Thus, the planar rattling mode of Cu2 is totally suppressed by the 
application of pressure. 

 The pressure effect on the planar rattling of Cu2 is very distinct from 
that on the rattling of Ba in a cage of BGS. The $C$/$T^{3}$ data of BGS at 
$P = 0$ show a broad maximum at 4 K, whose value agrees with that previously 
measured by PPMS (open circles) as shown in Fig. 2(a) \cite{MA05,KS10}. With 
increasing pressure up to 6.5 GPa, the broad maximum of $C$/$T^{3}$ 
monotonically shifts to higher temperatures, maintaining a maximum value 
at a level more than half for $P = 0$. This result indicates the survival 
of the rattling of Ba with an increase in the characteristic energy.

Next, we turn our attention to the upturn of $C$/$T^{3}$ observed at $T \, \textless \, 
1$ K for both CZSS and BGS. As shown in Fig. 1, the upturn for CZSS hardly 
changes with increasing pressure while that for BGS decreases drastically as 
shown in Fig. 2(a). This contrasting response implies distinct mechanisms of 
the upturn between the two systems. The upturn in BGS is due to the 
$T$-linear term of $C(T)$ which originates from the quantum tunneling of the guest 
atom among off-center potential minima in the oversized cage \cite{MA05,KS10,JW33}. 
Therefore, the reduction of $C$/$T^{3}$ at $T \, \textless \, 1$ K for BGS should be 
attributed to the pressure-induced change of the potential minimum from the 
off-center sites to the center in the cage. This interpretation is 
consistent with the Raman scattering study of rattling under pressure \cite{TS11} 
and the chemical pressure effect on the specific heat \cite{KS12,JW33}. It was found 
that the $T$-linear specific heat of BGS is suppressed when the Sn atoms on the 
cage are replaced by Ge atoms with a smaller ionic radius \cite{JW33}. On the 
contrary, the upturn of $C$/$T^{3}$ for CZSS remains unchanged even at $P \ge 
2.4$ GPa, where the specific heat due to the rattling of Cu2 atoms is 
completely suppressed as discussed above. Therefore, the rattling of Cu2 
atoms is not responsible for the upturn of $C$/$T^{3}$. We recall here that 
$C(T)$ for $T \, \textless \, 1$ K can be expressed by including a term proportional to 
$T^{-\alpha}$ ($\alpha \,  \textgreater \, 1$) \cite{KS31}. A similar behavior in 
cuprate superconductors was assigned to the Cu nuclear contribution as 
expressed by \textit{IT}$^{-2\, }$\cite{ND34}. Therefore, we fitted the specific heat data at 
$T \le  1$ K in terms of the form $C(T) =$ \textit{IT}$^{-2} + \gamma T + \beta 
T^{3}$, where $\gamma T$ and $\beta T^{3}$ represent the conduction-electron 
and Debye phonon contributions, respectively. Then, we calculated the phonon 
contribution $\Delta C$ as $\Delta C = C(T) -$ (\textit{IT}$^{-2} + \gamma T)$ for 
each pressure. 

As shown in Fig. 2(b), the maximum temperature $T_{\rm max}$ of $\Delta 
C/T^{3}$ for CZSS initially decreases from 4 K for $P =0$ to 3.4 K at 0.7 
GPa and slightly increases for $P \, \textgreater \, 1$ GPa, while the maximum of 
$C/T^{3}$ for BGS shifts to high temperatures monotonically as described above. 
In order to discuss quantitatively the pressure dependence of the 
vibrational density of states, we now analyze the specific heat data with 
the soft potential model (SPM) which could reproduce the broad maximum of 
$C/T^{3}$ for clathrates \cite{KS10,LG30,KU35}. This model assumes that the potential of 
the soft modes (SMs) for the rattler is given by $V(x) = W(D_{1}x +$ 
$D_{2}x^{2\, }+ x^{4})$, where $W$ is the characteristic energy of the 
potential, $x$ is the dimensionless displacement of the vibrating unit, and 
$D_{1}$ and $D_{2}$ are the coefficients of the asymmetry and 
harmonic-potential terms which vary from mode to mode \cite{MA36}. In addition to $W$, 
four parameters are used to fit the $C(T)$: the broadness of the SM, $A$; the distribution 
constant of the SM, $P_{\rm s}$; the typical experimental time; and the minimum 
relaxation time \cite{MA36}. This model assumes that the contributions from both the SM 
mode and tunneling mode (TM) of two-level systems increase at low 
temperatures \cite{MA36}. As shown by the solid line in Fig. 2(b), the broad 
maximum of the $\Delta C$/$T^{3}$ of CZSS at $P = 0$ was reproduced with the 
parameters $A = 0.045$, $W$/$k_{\rm B}= 4.0$ K, and $P_{\rm s}= 5.9 \times 
10^{21}$ mol$^{-1}$, without including the contribution of the TM. The data for 
BGS at $P = 0$ were fit with the parameters $A = 0.017$, $W$/$k_{\rm B}= 3.5$ K, and 
$P_{\rm s}= 6.7 \times 10^{21}$ mol$^{-1}$, which agree with previously 
reported values \cite{KS10}. The data of  $\Delta C$/$T^{3}$  of CZSS at low pressures and above 5 K could not be reproduced by the SPM. This discrepancy may be attributed to the fact that the phenomenological SPM model does not take into account the vibrational modes at a higher energy compared with the rattling energy.

Figures 2 (c) and 2(d) show, respectively, the vibrational density of states 
$g(\nu )$/$\nu^{2}$ for BGS and CZSS under various pressures as a function 
of the frequency $\nu $ used for the fit with the SPM model. The pressure 
dependences of $\nu_{\rm max}$ and the maximal values of $g(\nu )$/$\nu 
^{2}$ at $\nu = \nu_{\rm max}$ are displayed in Figs. 3(a)-(d). For 
BGS, the monotonic increase of $\nu_{\rm max}$ with increasing pressure 
suggests a gradual increment of harmonicity in the vibration of the Ba guest 
atom in the cage. A similar increase in the rattling energy for 
Sr$_{8}$Ga$_{16}$Ge$_{30}$ was observed when the cage was shrunk by the 
substitution of Si for Ge \cite{KS12}. For CZSS, the frequency at the peak $\nu 
_{\rm max} = 0.4$ THz corresponds to the rattling energy of 1.7 meV, which 
is between the two characteristic energies 1.0 and 2.8 meV obtained by an 
analysis of the specific heat with the Einstein model \cite{EL37}. The reduction in 
$\nu_{\rm max}$ up to 0.7 GPa for CZSS shown in Fig. 3(c) indicates the 
enhancement of anharmonicity in the vibration of the Cu2 atom. This result is 
reproduced by the soft-mode dynamic theory considering the interaction 
between the acoustic phonon and soft-mode vibrational excitations \cite{MB38}. The 
opposite $P$ dependences indicate that origins of anharmonicity are distinct 
between the out-of-plane rattling of Cu2 atoms in CZSS and the rattling of 
Ba ions in the cage of BGS. On further increasing pressure to $P \, \textgreater \,
1$ GPa, the value of $\nu_{\rm max}$ for CZSS increases steeply. This fact 
suggests the enhancement of harmonicity in the vibration of Cu2 atoms. It is 
noteworthy that the value of $g(\nu )$/$\nu^{2}$(max) for CZSS decreases 
with further increasing $P$ as shown in Fig. 3 (d). The additional contribution 
of SM over the Debye phonon contribution (dashed line) disappears for $P \, 
\textgreater \, 2.4$ GPa. The cessation of the vibrational density of states is in 
contrast with the case of BGS shown in Fig. 3(b), where $g(\nu )$/$\nu 
^{2}$(max) levels off at around 4 GPa but remains at half of the initial 
value even at 6 GPa. The very different $P$ dependence indicates that the 
planar rattling of Cu2 atoms in CZSS is more fragile to pressure than the 
rattling of Ba atoms in the cage of BGS. 

Table I shows the structural parameters for CZSS under pressures of 0.3 and 
1.9 GPa at room temperature. We analyzed the powder x-ray diffraction 
patterns by using an on-site model and a split-site model. The on-site 
model assumes that atoms are located at the crystallographic sites in the 
cubic structure ($I\bar{4}3m$): Cu1 at 12$d$ (1/2,0,1/4), Cu2 at 12$e$ (0,0,$z)$, Sb at 
8$c$ ($x$,$x$,$x)$, S1 at 24$g$ ($x$,$x$,$z)$, and S2 at 2$a$ (0,0,0) \cite{KS17,YB19,AP20,NG25,VR26,SO27,KS28,WL29}. In the 
split-site model, the site for Cu2 splits into a 24$g$ site (Cu22) out of the 
S$_{3}$ triangle and the original 12$e$ site (Cu21) \cite{AP20}, and 
simultaneously the site of S2 splits into a 12$e$ site (S22) and the original 
2$a$ site (S21). Occupancies and shifts of additional sites in the split-site 
model were almost zero in the refinement of the 0.3 GPa diffraction pattern. 
The x-ray diffraction pattern was well reproduced by the on-site model. This 
result is consistent with previous refinements of the diffraction data 
at ambient pressure using the two models \cite{VR26,KS28}. On the contrary, the 
Rietveld refinement of the pattern at 1.9 GPa with the split-site model gave 
reliability factors of Bragg intensities, $R_{\rm I\, }= 5.088${\%}, and the  
weighted profile, $R_{\rm wp} = 1.282${\%}, whose values are, respectively, 
smaller than those of the refinement with the on-site model ($R_{\rm I}= 
5.224${\%} , $R_{\rm wp} = 1.338${\%}). This means that the Cu2 atom at 1.9 
GPa has a large probability to be located at the local potential minimum out 
of the S$_{3}$ triangle. 

Based on the structural parameters for CZSS, we discuss the change in the 
rattling energy of the Cu2 atom upon applying pressure up to 0.7 GPa. As shown 
in Fig. 3(c), the rattling energy of $\nu_{\rm max}$ for CZSS decreases by 
14{\%} from 0.37 THz at 0 GPa to 0.32 THz at 0.7 GPa. At first, we focus on 
the role of lone pairs of Sb for the rattling of the Cu2 atom. Lai \textit{et al}. claimed 
that covalent-type ``out-of-plane bonding'' between Cu2 and Sb via the lone 
pairs leads to a rattling of Cu2 \cite{WL29}. The Cu2-Sb distance, $d$(Cu2-Sb), is a 
measure of the free space for the out-of-plane vibration of Cu2 as well as 
that of the magnitude of electron sharing between Cu2 and Sb. The decreasing 
ratio of $d$(Cu2, Cu21-Sb) upon pressurizing from 0.3 to 1.9 GPa is estimated 
to be 0.029 {\AA}/GPa using the values listed in Table I. This ratio gives 
a reduction in $d$(Cu2-Sb) of 0.02 {\AA} upon pressurizing from $P = 0$ to 0.7 
GPa. This value is only 0.6{\%} for $d$(Cu21-Sb) $=$ 3.344 {\AA} \cite{KS28}. The 
small change in $d$(Cu2-Sb) suggests a weak and positive correlation between 
$\nu_{\rm max}$ and $d$(Cu2-Sb). On the contrary, a negative correlation was 
reported for the substituted system of CSS \cite{KS28}. In fact, the rattling 
energy $E_{\rm R}$ of Cu2 decreases by 19{\%} ($\nu_{\rm max}$ by 14{\%}) by 
the substitution of As for Sb in CSS \cite{KS28}, where $d$(Cu2-As) is elongated 
compared to $d$(Cu2-Sb). These comparisons of the pressure effect with the 
substitution effect indicate the minor role of lone pairs in the rattling of the  
Cu2 atom.

In turn, we discuss the relationship between $E_{\rm R}$ of the Cu2 atom and the 
S$_{3}$ triangle area $S_{\rm tri}$. It was reported that $E_{\rm R}$ decreases with 
decreasing $S_{\rm tri}$ on going from CZSS, CSS, 
Cu$_{10}$Zn$_{2}$As$_{4}$S$_{13}$, to Cu$_{12}$As$_{4}$S$_{13}$ \cite{KS28}. Let us 
compare the pressure effect on $E_{\rm R}$ with the substitution effect. As shown 
in Table I, the average value of 6.07(3) {\AA}$^{2}$ of $S_{\rm tri}$ at $P = 1.9$ 
GPa is 3.5{\%} smaller than that for $P = 0.3$ GPa. Therefore, the decreasing 
ratio of $S_{\rm tri}$ is 2.2 {\%}/GPa. From Fig. 3(c), the decreasing ratio of 
the rattling energy of $\nu_{\rm max}$ is 20{\%}/GPa in the $P$ range from 0 to 
0.7 GPa. These values yield $d$(ln$\nu_{\rm max})$/$d$(ln$S_{\rm tri}) = 9$, which 
agrees with $d$(ln$E_{\rm R})$/$d$(ln$S_{\rm tri})$. The latter was estimated from the 
analysis for the substituted systems CZSS, CSS, 
Cu$_{10}$Zn$_{2}$As$_{4}$S$_{13}$, and Cu$_{12}$As$_{4}$S$_{13}$ \cite{KS28}. This 
agreement indicates that $S_{\rm tri}$ is the dominant parameter that controls 
the rattling of the Cu2 atom in CZSS. Thus, it is proved that the rattling in 
CZSS originates from the chemical pressure inherent in the S$_{3}$ triangle 
to squeeze the Cu2 atom out of the plane. 

We now turn to a discussion of the vibrational state of the Cu2 atom in CZSS for the 
range $P \, \textgreater \,  1$ GPa. As shown in Figs. 2(d) and 3(d), the density of states for the 
rattling of the Cu2 atom decreases strongly for $P \, \textgreater \, 1$ GPa and 
disappears for $P \, \textgreater \, 2.4$ GPa. Furthermore, as presented in Fig. 3(c), 
the rattling energy $\nu_{\rm max}$ increases for $P \, \textgreater \, 1$ GPa, 
indicative of the enhancement of harmonicity in the vibration of the Cu2 atom. 
As shown in Table I, the distance between Cu22 and Sb is 2.49(6) {\AA} at 
1.9 GPa, which is close to the distance of 2.2 {\AA} between Cu21 and S. 
Thereby, the bond order value of about 0.5 for the Cu22-Sb pair becomes five 
times larger than that for the Cu21-Sb pair reported in Ref. 28. These facts 
allow us to speculate that the Cu2 atom is ejected from the S$_{3}$ plane due to 
the chemical pressure, and is combined with Sb for $P \, \textgreater \, 2$ GPa. This 
scenario is consistent with the suppression of the vibrational density of 
states for the rattling of Cu2 atoms. 

In summary, we have studied the pressure effect on the rattling mode in 
tetrahedrite Cu$_{10}$Zn$_{2}$Sb$_{4}$S$_{13}$ (CZSS) and type-I clathrate 
Ba$_{8}$Ga$_{16}$Sn$_{30}$ (BGS) by specific heat measurements up to 6 
GPa. For $P = 0$, the $C$/$T^{3}$ data for both systems exhibit a maximum at 4 K, 
five times of which is a measure of the rattling energy. With increasing 
pressure, the energy for CZSS initially decreases and steeply increases for 
$P \, \textgreater \, 1$ GPa while that for BGS increases monotonically. Furthermore, 
the upturn of $C$/$T^{3}$ at $T \, \textless \,1$ K for CZSS hardly changes with 
increasing pressure while that for BGS decreases drastically. These 
contrasting responses of $C$/$T^{3}$ are attributed to distinct mechanisms for 
the rattling between the two systems. For the Ba guest ions in an oversized 
cage of BGS, an application of pressure shifts the potential minimum from the  
off-center sites to the center in the cage, and thus increases the harmonicity 
in the vibration of Ba ions. 

For CZSS, the synchrotron powder x-ray diffraction measurement has revealed 
that the area of S$_{3}$ triangle, the $S_{\rm tri}$, decreases with increasing 
pressure up to 1.9 GPa. The systematic dependences of the rattling energy on 
$S_{\rm tri}$ by an application of pressures and substitutions strongly indicate 
that the dominant parameter controlling the rattling of the Cu2 atom in CZSS is 
not the distance $d$(Cu2-Sb) but the area $S_{\rm tri}$. These findings support that 
the rattling in CZSS originates from the chemical pressure inherent in the 
triangle to squeeze the Cu2 atom out of the plane. On further increasing 
pressure up to 2.4 GPa, the specific heat due to the rattling modes 
completely disappears. This is consistent with the splitting of the Cu2 site 
as indicated by the x-ray diffraction analysis on the data at 1.9 GPa. At 
$P \, \textgreater \, 2$ GPa, the Cu2 atom is likely to vibrate near the local 
potential minimum out of the S$_{3}$ triangle. 

Our study verified high-pressure specific heat measurement as a strong tool 
to investigate the rattling modes in solids. Furthermore, the pressure-sensitive 
planar rattling modes found in tetrahedrite would allow us to provide different 
guidelines for the development of high-performance thermoelectric materials.

\begin{acknowledgments}

We acknowledge valuable discussions with C. H. Lee and T. Onimaru. The 
specific heat measurement under pressures were performed at N-BARD, Hiroshima 
University. This work was partly supported by Japan Society for the 
Promotion of Science KAKENHI Grants No. JP25400375, No. JP16H01073, No. JP18K03518, No. JP18H04324, No. JP18H04499, and CREST JST 
Grants No. JPMJCR16Q6, and No. JPMJCR20Q4. Synchrotron radiation x-ray 
diffraction experiments were carried out at SPring-8 with the approval of 
the Japan Synchrotron Radiation Research Institute (Proposals No. 2018B0078 and No. 2019A0159).

\end{acknowledgments}

\bibliography{basename of .bib file}

\newpage

%



\begin{figure}
\includegraphics[width=10cm]{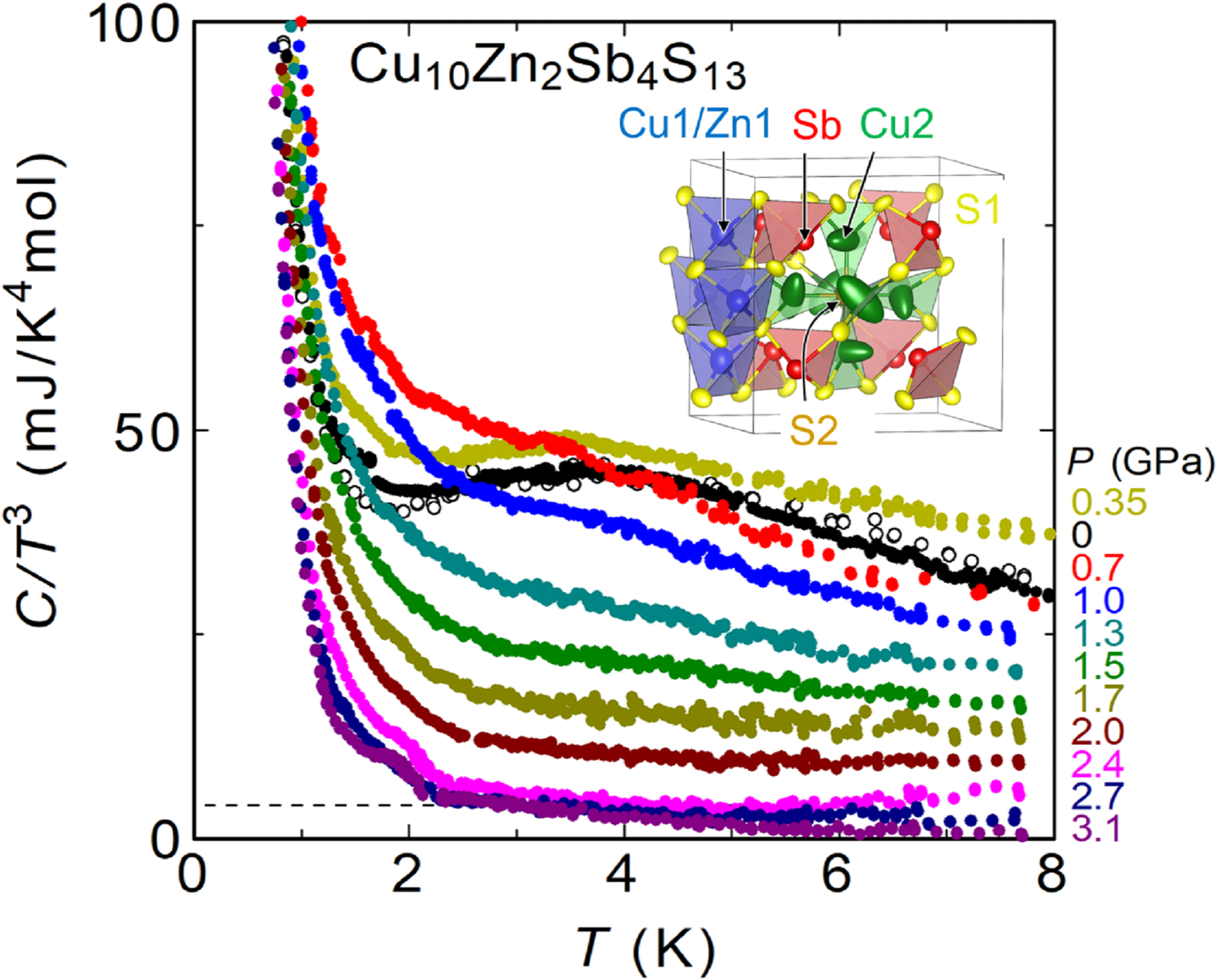}%
\caption{\label{fig:epsart}Temperature dependence of the specific heat $C$ divided by $T^{3}$, 
$C$/$T^{3}$, for Cu$_{10}$Zn$_{2}$Sb$_{4}$S$_{13}$ under various constant 
pressures up to 3.1 GPa. The data for $P = 0$ (solid black circles) agree 
with the previous data (open black circles) measured by Quantum Design PPMS 
\cite{KS31}. The dashed line at the bottom represents the calculation using the 
Debye model with a Debye temperature of 242 K \cite{KS31}.
}
\end{figure}

\newpage

\begin{figure}
\includegraphics[width=10cm]{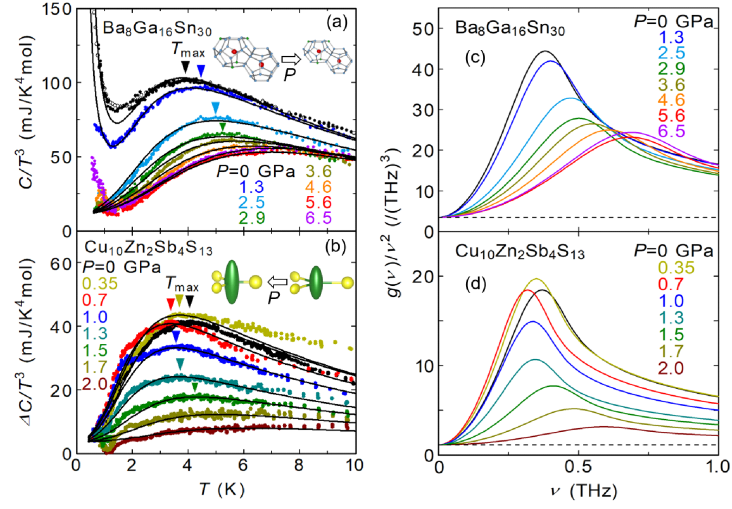}%
\caption{\label{fig:epsart} 
(a) Specific heat $C$ divided by $T^{3}$, $C$/$T^{3}$ vs $T$ for 
Ba$_{8}$Ga$_{16}$Sn$_{30\, }$(BGS) up to 6.5 GPa, and (b) the lattice contribution 
of $C$/$T^{3}$, $\Delta C$/$T^{3}$, vs $T$ for Cu$_{10}$Zn$_{2}$Sb$_{4}$S$_{13\, }$
(CZSS) under pressures up to 2 GPa. Solid lines are fits using the soft 
potential model (SPM). The shoulders in $\Delta C$/$T^{3}$ near 2 K for CZSS are experimental artifacts caused by no smooth connection of calibration curves of the thermometer used for the specific heat measurements. The vibrational density of states $g(\nu )$/$\nu 
^{2}$ calculated by SPM for (c) BGS and (d) CZSS as a function of 
frequency $\nu $. The dashed lines represent calculations using the Debye 
model with Debye temperatures of 210 K \cite{KS10} and 242 K \cite{KS31} for BGS and 
CZSS, respectively.
}
\end{figure}

\newpage

\begin{figure}
\includegraphics[width=10cm]{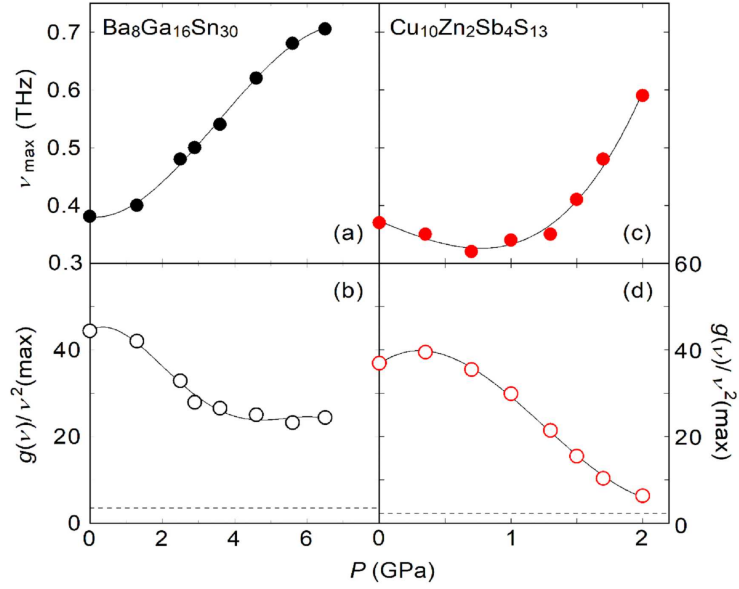}%
\caption{\label{fig:epsart}
Pressure dependences of the frequency $\nu_{\rm max}$ at which the 
vibrational density of states $g(\nu )$/$\nu^{2}$ in Figs. 2(c) and 2(d) 
has the maximum, and a maximal value of $g(\nu )$/$\nu^{2}$ for 
Ba$_{8}$Ga$_{16}$Sn$_{30}$ [(a), (b)] and Cu$_{10}$Zn$_{2}$Sb$_{4}$S$_{13}$ 
[(c), (d)]. Solid lines are guides for the eyes. The dashed lines in (b) and 
(d) represent calculations using the Debye model with Debye temperatures 
of 210 K \cite{KS10} and 242 K \cite{KS31} for BGS and CZSS, respectively.
}
\end{figure}

\newpage

\begin{table*}
\caption{\label{tab:table3}
Crystallographic parameters of Cu$_{10}$Zn$_{2}$Sb$_{4}$S$_{13}$ 
for $P = 0.3$ and 1.9 GPa at 300 K: Lattice constant, atomic coordination, 
occupancy, interatomic distances, and area of S$_{3}$ triangle, $S_{\rm tri}$.
}
\begin{ruledtabular}
\begin{tabular}{lll}
 
Parameter& 
0.3 GPa (onsite)& 
1.9 GPa (split site) \\
\hline
$a$ ({\AA})& 
10.3429 (3)& 
10.2607 (3) \\

$R_{\rm wp}$ ({\%}) & 
1.142& 
1.282 \\

$R_{\rm I}$ ({\%}) & 
3.769& 
5.088 \\

Cu1 in 12$d $(1/2, 0, 1/4); occ. & 
2/3& 
2/3 \\

Zn1 in 12$d$ (1/2, 0, 1/4); occ.& 
1/3& 
1/3 \\

Cu2 in 12$e$ (0, 0, $z)$; occ.& 
$z=0.2167$; 1.0& 
$-$ \\

Cu21 in 12$e$ (0, 0, $z)$; occ. &
$-$& 
$z=0.2172$ (6); 0.85 \\

\begin{tabular}{l}
Cu22 in 24$g$ ($x$, $-x$, $z)$; occ. \\ \ 
\end{tabular}&
\begin{tabular}{l}
$-$ \\ \ 
\end{tabular}&
\begin{tabular}{l}
$x=0.061$ (5), $z=0.191$ (4);\\0.075
\end{tabular}\\

\begin{tabular}{l}
S1 in 24$g$ ($x$, $x$, $z)$; occ. \\ \ 
\end{tabular}& 
\begin{tabular}{l}
$x=0.1141$ (5), $z=0.3644$ (4);\\1.0
\end{tabular} & 
\begin{tabular}{l}
$x=0.1115$ (7), $z=0.3646$ (5);\\1.0
\end{tabular}\\

S2 in 2$a$ (0, 0, 0); occ.&
1.0& 
$-$ \\

S21 in 2$a$ (0, 0, 0); occ. &
$-$& 
0.15 \\

S22 in 12$e$ ($x$, 0, 0); occ.& 
$-$& 
$x=0.058$ (3); 0.1417 \\

Sb in 8$c$ ($x$, $x$, $x)$; occ.& 
$x=0.2684$ (3); 1.0& 
$x=0.2697$ (4); 1.0 \\

$d$(Cu2-Sb) ({\AA}) & 
3.391 (3)& 
$-$ \\

$d$(Cu21-Sb) ({\AA}) & 
$-$& 
3.344 (5) \\

$d$(Cu22-Sb) ({\AA}) & 
$-$& 
2.49 (6) \\

$d$(S1-S1) ({\AA}) & 
3.337 (11)& 
3.235 (15) \\

$d$(S1-S2) ({\AA})& 
4.121 (5)& 
$-$ \\

\begin{tabular}{l}
$d$(S1-S21), $d$(S1-S22) ({\AA})  \\ \ 
\end{tabular}& 
\begin{tabular}{l}
$-$ \\ \ 
\end{tabular}& 
\begin{tabular}{l}
4.075 (6), 3.54 (4), 3.950 (8), \\4.279 (16), 4.62 (4) \\
\end{tabular}\\

\begin{tabular}{l}
$S_{\rm tri}$ ({\AA}$^{2})$ \\ \ 
\end{tabular}& 
\begin{tabular}{l}
6.29 (3) \\ \ 
\end{tabular}& 
\begin{tabular}{l}
6.05 (4), 5.09 (12), 6.09 (5), \\7.00 (16); \par Ave., 6.07 (3) \\
\end{tabular}\\

\end{tabular}
\end{ruledtabular}
\end{table*}

\end{document}